\documentclass[english,english,aps,prd,amsmath,amssymb,superscriptaddress]{revtex4}
\usepackage[T1]{fontenc}
\usepackage[latin9]{inputenc}
\usepackage{amsmath}
\usepackage{amssymb}
\usepackage{esint}

\makeatletter
\@ifundefined{textcolor}{}
{%
 \definecolor{BLACK}{gray}{0}
 \definecolor{WHITE}{gray}{1}
 \definecolor{RED}{rgb}{1,0,0}
 \definecolor{GREEN}{rgb}{0,1,0}
 \definecolor{BLUE}{rgb}{0,0,1}
 \definecolor{CYAN}{cmyk}{1,0,0,0}
 \definecolor{MAGENTA}{cmyk}{0,1,0,0}
 \definecolor{YELLOW}{cmyk}{0,0,1,0}
 }

\usepackage{slashed}
\usepackage{babel}
\usepackage{bbm}

\makeatother

\usepackage{babel}

\begin{document}

\title{Dynamic Critical Scaling of the Holographic Spin Fluctuations}

\author{M.J.Luo}

\affiliation{Department of Modern Physics, University of Science and Technology
of China, Anhui 230026, People's Republic of China}
\begin{abstract}
Criticality with strong coupling is described by a theory in the vicinity
of a non-Gaussian fixed point. The holographic duality conjectures
that a theory at a non-Gaussian fixed point with strong coupling is
dual to a gravitational theory. In this paper, we present a holographic
theory in treating the strongly coupled critical spin fluctuations
in quasi-2-dimension. We show that a universal frequency over temperature
scaling law is a rather general property of the critical ac spin susceptibility
at strongly coupled limit. Explicit results for the dynamic scaling
of spin susceptibility are obtained in large-N and large 't Hooft
limit. We argue that such critical scaling are in good agreement with
a number of experiments, some of which can not be explained by any
perturbative spin-density-wave theory. Our results strongly suggest
that the anomalous behavior of non-Fermi liquids in materials is closely
related to the spin fluctuations described through the non-Gaussian
fixed point. The exotic properties of non-Fermi liquids can be viewed
as the Fermi liquids coupling to strongly coupled critical spin fluctuations.
\end{abstract}
\maketitle

\section{Introduction}

The subjets of spin fluctuations have attracted wide attentions \cite{RevModPhys.63.1,2010Natur.464..199B},
due to the intense interests in understanding the exotic properties
of the heavy fermion (HF) metals and the normal state of non-conventional
superconductors. The HF metals are materials in which the effective
mass of the charge carriers is often hundreds or even thousands of
times greater than the mass of bare electrons. It is now clear that
the large effective mass arises from the hybridization between the
conducting electrons and the magnetic moments or spins of the parent
compounds (Kondo effect) \cite{2006cond.mat.12006C}. The phase diagram
of cuprate high-Tc superconductors looks similar to that of the HF
metals, except that there is a pseudogap regime separating a weakly
doped anti-ferromagnetic and superconducting phase around optimal
doping. Spin fluctuations play an important role not only in its anti-ferromagnetic
regimes, but also in the superconducting dome, qualitatively, d-wave
Cooper pairs can be formed by exchanging spin fluctuations between
electrons\cite{PhysRevB.34.8190}.

It is by now established that the parent compounds of these two materials,
at least at weakly doping, are well described through spin dynamics
by using a spin-1/2 lattice Heisenberg model with nearest-neighbor
interaction. As gradually chemical doped, such materials carrying
constituent magnetic moments/spin fluctuations have been observed
to develop quantum criticality when their transition temperature is
driven to nearly zero \cite{2008NatPh...4..186G}, known as the Quantum
Critical Point (QCP). The spin degrees of freedoms are present at
all temperatures down to the QCP \cite{2000Natur.407..351S}. 

As the spin fluctuations become critical, the system exhibits strongly
coupled dynamics due to the destruction of Kondo screening by the
critical fluctuations. In this case, the critical spin flutuation
can not trivially explained by a weakly coupled spin-density-wave
\cite{2000Natur.407..351S} . The properties of the material dramatically
change by the critical fluctuations and form the so-called non-Fermi
liquids or {}``strange metal'' phase, the anomalous behaviors of
which depart the standard Landau's Fermi-liquid theory that successfully
describing most metals, for a review see \cite{2008NatPh...4..173S}.
The non-Fermi liquid behavior is closely related to the strongly coupled
new state of matter in the Quantum Critical Regime (QCR), which is
the finite temperature extension of a QCP. Understanding the behaviors
of non-Fermi liquids is a major challenge in condensed matter physics.
However, the mechanisms are still highly controversial, being almostly
due to the lack of exact theoretical computations on such strongly
coupled critical systems.

The strongly coupled critical spin fluctuation as an example of quantum
critical states is a tremendous difficult problem ever studied, see
e.g. \cite{PhysRevB.49.11919}, and an important goal of theoretical
studies in understanding non-Fermi liquids. There are trials in treating
the system preserving traditional notion of quasi-particles, e.g.
the Fermion condensation quantum phase transition theory, see review
\cite{2010PhR...492...31S}. However, restrictly speaking, there is
no well-defined quasi-particles/waves any more and even no Landau's
local order parameter description in most cases \cite{2001AdPhy..50..361L,2003Natur.424..524C,2011PhT....64b..29S},
which makes the system completely beyond the scope of using traditional
method and needs completely new idea. 

If one notes that near the critical point the correlation length is
much larger than the microscopic length scale, the details of its
microscopic structures is unimportant, so the fluctuations near the
criticality behaves universal, in addition to the strongly coupled
dynamics, it is more promising to conjecture that the strongly coupled
critical system should be described by AdS/CFT correspondence or holographic
duality discovered in string theory \cite{Maldacena:1997re,Witten:1998qj}.
There are examples and evidences show that the holographic approach
is much heuristic and even trustable, one successful prediction \cite{Policastro:2001yc,Policastro:2002se}
was that the ratio of the shear viscosity to the entropy density of
the strongly coupled quark-gluon plasma near the critical transition
temperature, which agrees well with the measurements of it produced
from relativistic heavy ion collision \cite{PhysRevLett.97.162302}.
Different from the conventional renormalization group analysis of
critical behavior, which relies largely on the existence of a Gaussian
fixed point and the validity of perturabative calculation near the
fixed point. The strongly coupled Conformal Field Theory (CFT) without
doubt describes a non-Gaussian fixed point or critical point of the
system beyond the perturbation theory. Our discussion of the critical
dynamic scaling will be based on such non-Gaussian fixed point, which
are suggested in this paper response for the anamolous behavior of
non-Fermi liquids.

This holographic approach has been actively applied to wide varieties
of condensed matter problems in recent years, see reviews \cite{2009CQGra..26v4002H,2010arXiv1002.2947S},
e.g. holographic superconductor \cite{Hartnoll:2008vx,Hartnoll:2008kx}
and superfluidity \cite{Herzog:2008he,Herzog:2009xv}, holographic
strange metals \cite{Faulkner:2010da,Sachdev:2010uj,2010Sci...329.1043F}
and non-Fermi liquid \cite{Lee:2008xf,Liu:2009dm,Faulkner:2011tm},
semi-holographic Fermi liquid \cite{Faulkner:2010tq,Sachdev:2010um},
and quantum criticality \cite{Sachdev:2008ba,Faulkner:2009wj,Iqbal:2010eh,Iqbal:2011ae,Iqbal:2011aj}.
The holographic superconductor or superfluidity implements a second
order phase transition or mildly crossover by condensation of strongly
coupled charged scalar fields (and more complicated higher integer
spin fields generalizations \cite{Gubser:2008wv,Chen:2010mk}) because
of the intrinsic holographic instability \cite{Gubser:2008px}. In
the works on holographic strange metals, strongly coupled Fermionic
fields with spin-1/2 are studied by holographic approach, which realizes
a non-Fermi liquid behavior with linear temperature resistivity. The
semi-holographic Fermi liquid theory simplifies and generalizes the
idea from holographic strange metals, introducing a hybridization
between free fields and strongly coupled critical modes at infrared
described by $\mathrm{AdS_{2}}$ near the horizon of blackhole.

The purpose of the paper is to examine the critical dynamic scaling
behavior of the correlation function of strongly coupled critical
spin fluctuations by holographic approach, the calculations lead to
variety of testable predictions, and we try to connect our calculations
to experimental observations at least qualitatively. A discussion
for the correlation function of dual currents in 2+1 dimensional can
be found in \cite{Herzog:2007ij}. However, in this paper, we find
the 3+1 dimensional holographic model is a more realistic model, because
(i) the 3+1D holographic spin system is naturally a quasi-2-dimensional
(quasi-2D) system similar to the real 2D-layer of a compound, as a
consequence, the spin susceptibility is only a scaling of frequency
over temperature ($\omega/T$) independent to wave vector in the quasi-2D
plane; (ii) the scaling of spin ac susceptibility $\chi\sim(\omega/T)^{\alpha}$
takes a value $\alpha=1$ at low frequency hydrodynamic limit, and
$\alpha=2/3$ at high frequency relativistic limit which can not be
explained by any spin-density-wave QCP theory; (iv) the dc spin susceptibility
approaches a constant independent to temperature; (v) these scaling
behaviors are very crucial in understanding the exotic properties
of non-Fermi liquid, and we show that our results are in good agreement
with a number of experiments. The strongly coupled critical spin fluctuations
in our paper play similar role with the infrared critical modes discussed
in the semi-holographic Fermi liquid theory \cite{Faulkner:2010tq,Sachdev:2010um}.
The bulk of the paper is devoted to making these statements more precise.

The paper is organized as follows. We present the general arguments
of holographic spin dynamics in Sec.\ref{sec:Holographic-Spin-Dynamics}.
In Sec.\ref{sec:Scaling}, we formulate the technical detail of the
approach and evaluate the scaling exponent of spin susceptibility
in three analytically tracetable limits. In Sec.\ref{sec:Connection-to-experiments},
we discuss the comparison of our results with experimental phenomenon
in two aspects, including qualitatively explaining the scaling law
of spin susceptibility at criticality and linear-temperature resistivity
in strange metals. Finally, we conclude in Sec.\ref{sec:Conclusions}.

\section{\label{sec:Holographic-Spin-Dynamics}Holographic Spin Dynamics}

\subsection{Non-Gaussian Fixed Point}

A theoretical model for considering the spin dynamics doped with conduction
electrons is described by the Kondo lattice model,

\begin{equation}
\mathcal{H}=\sum_{\langle i,j\rangle,s}t_{ij}c_{is}^{\dagger}c_{js}+J_{K}\sum_{i,s,s^{\prime}}c_{is}^{\dagger}c_{is^{\prime}}\mathbf{\sigma}_{ss^{\prime}}\cdot\mathbf{S}_{i}+\sum_{\langle i,j\rangle}I_{ij}\mathbf{S}_{i}\cdot\mathbf{S}_{j},\label{eq:Kondo_lattice_model}\end{equation}
where $c_{i}$ is the electron operator and $\mathbf{S}_{i}$ is the
on site spin operator at lattice site $i$. The first term is the
hopping of the conduction electrons between neighbor site i and j.
At each lattice site, the local spin $\mathbf{S}_{i}$ interacts via
an exchange coupling $J_{K}$ with the spin of conduction electrons
sitting at the site. The local spin $\mathbf{S}_{i}$ interacts with
the nearest neighbor spin $\mathbf{S}_{j}$ by the Ruderman-Kittel-Kasuya-Yosida
(RKKY) via exchange interaction $I_{ij}$. 

The effective Hamiltonian for the on site local spin can be obtained
by integrating out the fermionic degrees of freedom, i.e. the conduction
electrons $c_{i}^{\dagger},c_{i}$, and then all the effects of the
conduction electron are absorbed into the the new effective exchange
interaction $J_{ij}$, we get the effective spin dynamics described
by the Heisenberg Hamiltonian, \begin{equation}
\mathcal{H}_{eff}=\sum_{\left\langle ij\right\rangle }J_{ij}\mathbf{S}_{i}\cdot\mathbf{S}_{j}.\label{eq:Heisenberg_model}\end{equation}

The action of such effective Heisenberg model in the vicinity of critical
point can be reduced to an equivalent continuum field theory, written
in terms of stagger order parameter, \begin{equation}
S_{0}[\mathbf{n}]=\int d\tau d\mathbf{r}\left[\left(\nabla_{\mathbf{r}}\mathbf{n}\right)^{2}+\left(\partial_{\tau}\mathbf{n}\right)^{2}+a\mathbf{n}^{2}+b\left(\mathbf{n}^{2}\right)^{2}\right],\label{eq:action_n}\end{equation}
in which 

\begin{equation}
\mathbf{n}(\mathbf{r}_{i})=e^{i\mathbf{Q}\cdot\mathbf{r}_{i}}\mathbf{S}_{i}\label{eq:order_spin}\end{equation}
is the anti-ferromagnetic order parameter and $\mathbf{Q}$ is wavevector
with each component at $Q=\pi$. 

The criticality of the model in Eq.(\ref{eq:Kondo_lattice_model})
or Eq.(\ref{eq:action_n}) could develop via the competition between
the fermionic and bosonic environments interacting with the local
spin: the quenching of the local spin through its Kondo coupling to
the conduction electron bath and the coupling between local spin and
fluctuating magnetic field generated by the local spins at all other
sites. At the QCP, the effects of the fermionic and bosonic baths
come into balance, and the bosonic coupling succeeds in preventing
the local spins from completely quenching, leading to a singular spin
susceptibility. 

The mechanism that develops criticality reflecting in the effective
action Eq.(\ref{eq:action_n}) can be manifested in a more transparent
way. Since the fermionic bath have been integrated out in the effective
action, the criticality appears directly from the bosonic action.
A rough observation in Eq.(\ref{eq:action_n}) shows that there exists
two types of fundamentally different fixed points or critical points
in the system. The first one is trivial, when $a\rightarrow0,b\rightarrow0$,
which represents a trivial Gaussian fixed point without interaction,
corresponding to the quenched free spin wave. It is worth noting that
the Wilson-Fisher fixed point found at low dimension (d<4) also resides
in its perturbative character which is applicable only when the coupling
$b$ is sufficiently small, and hence the spin fluctuations also behave
as quasi-particles. However, according to the AdS/CFT duality, it
is conjectured that there exists a non-Gaussian fixed point when the
interaction strength $b$ is non-zero and takes large value. It corresponds
to a strong tendency to polarize the local spin along the direction
of the fluctuating magnetic field generated by the effective spin
environment and hence represents a strong self-coupling between spin
fluctuations. A comprehensive interpretation is as follows: the coupling
with fermionic bath leads to a negative contribution to the typical
energy scale characterized by parameter $a$, while the bosonic coupling
gives it a positive contribution, so when the coupling takes certain
large value, the typical energy scale vanishes and the system manifests
as an interaction criticality.

\subsection{Holographic Description of the Non-Gaussian Fixed Point}

These two types of fixed points relating to weak and strong coupling
strength belong to different universality classes with different critical
exponents. The dynamic critical exponents of the strongly coupled
non-Gaussian fixed point is the major interest of the paper, since
what we concern in this paper is the materials in the QCR exhibiting
strongly coupled dynamics.

In the vicinity of the strongly coupled non-Gaussian fixed point,
the theory is beyond the standard weak coupling technique and notoriously
difficult to solve, the action Eq.(\ref{eq:action_n}) is then no
longer directly useful for analytic perturbative calculations. However,
the critical nature of the strongly coupled non-Gaussian fixed point
indicates that it is a strongly coupled Conformal Fields Theory (CFT)
which is conjectured to be dual to a string/M theory in an AdS space
at large spin component limit. That is to say, the theory described
by the holographic approach is a theory in the vicinity of the non-Gaussian
fixed point. As a consequence, there must exists a holographic description
to the large-N critical spin system at strongly coupled limit.

The most fundamental question of the critical spin fluctuation concerns
the two-point retarded correlation function of spins, i.e. the spin
susceptibility, which measures in response to weak applied external
magnetic fields. It requires introduction of weak external magnetic
fields $\mathfrak{B}$ into the model by coupling the spin to the
$\mathfrak{B}$ fields \begin{equation}
S=S_{0}+\int d\tau d\mathbf{r}\mathbf{S}\cdot\mathfrak{B}.\end{equation}
In this paper, we will study the spin susceptibility defined by \begin{align}
\chi_{ij}(\omega,\mathbf{q}) & =-i\int d\mathbf{r}dt\theta(t)\langle[S_{i}(t,\mathbf{r}),S_{j}(0,0)]\rangle e^{-i\omega t+i\mathbf{q}\cdot\mathbf{r}}.\label{eq:definition}\end{align}
Obviously, if the system is anti-ferromagnetic as the action Eq.(\ref{eq:action_n})
refers to, the correlation function of the order parameter $\mathbf{n}$
must be anti-ferromagnetic. We can also define a spin susceptibility
by the anti-ferromagnetic order parameter. By using the Eq.(\ref{eq:order_spin}),
we have the relation between our definition Eq.(\ref{eq:definition})
and the correlator in terms of the order parameters by\begin{align}
\chi_{ij}(\omega,\mathbf{q})\delta(\mathbf{Q}) & =-i\int d\mathbf{r}dt\theta(t)\langle[n_{i}(t,\mathbf{r}),n_{j}(0,0)]\rangle e^{-i\omega t+i\mathbf{q}\cdot\mathbf{r}}.\end{align}
in which the delta-function peaking at $\mathbf{Q}$ exhibits the
expectation value is taken in an anti-ferromagnetic system.

The susceptibility defined in Eq.(\ref{eq:definition}) can be directly
obtained by evaluating the functional derivative of the partition
function with respect to the external weak magnetic fields, according
to the linear response theory,\begin{equation}
\chi_{ij}=\frac{2}{V}\frac{\delta^{2}\ln Z}{\delta\mathfrak{B}_{i}\delta\mathfrak{B}_{j}},\label{eq:chi_from_z}\end{equation}
where $V$ is the volume of the spin system. Applying the AdS/CFT
duality, the partition function of the strongly coupled critical spin
fluctuations can be calculated from the gravitational side, in large-N
and large 't Hooft coupling limit the gravitational theory reduces
to a classical gravity, the classical action gives a dominant contribution
to the partition function by the saddle point approximation,\begin{equation}
Z[B]=\langle e^{-\int_{\partial\mathcal{M}}\mathbf{S\cdot B_{0}}}\rangle_{NG}=e^{-S_{cl}[\mathbf{B}]},\label{eq:z}\end{equation}
where the subscript {}``NG'' denotes the expectation value is taken
at the non-Gaussian strongly coupled fixed point, $\partial\mathcal{M}$
stands for the boundary of the AdS space, i.e. our flat spacetime,
$\mathbf{B_{0}}$ is the boundary value of the bulk magnetic fields
$\mathbf{B}$. $S_{cl}$ is the classical action for the magnetic
fields propagating in an asymptotic AdS metric,\begin{equation}
S_{cl}[F]=-\frac{1}{4g_{YM}^{2}}\int d^{d+1}x\sqrt{-g}F_{IJ}F^{IJ},\label{eq:action_f}\end{equation}
where $F_{IJ}$ is the field strength, and $g_{YM}^{2}=16\pi^{2}R/N^{2}$
the gauge coupling. So we have holographic version of spin susceptibility\begin{equation}
\chi_{ij}=\frac{2}{V}\frac{\delta^{2}S_{cl}}{\delta B_{0i}\delta B_{0j}}.\label{eq:chi_from_S}\end{equation}

The strongly coupled critical spin fluctuations lived on the boundary
of the AdS space share features of the critical theory of physical
interest, there are no well-defined notion of quasi-particle, it is
insensitive to microscopic details and displays universal behavior,
the surprising success of the holographic approach at low energies
also exhibit universality in its predictions. The holographic spin
system captures the key features of the strongly coupled critical
spin fluctuations. Historically, scaling plays a central role in the
studying of criticality, so in the following section, we will turn
to the calculation of the scaling behavior of the holographic spin
fluctuations.

\section{\label{sec:Scaling}Scaling of the Spin Susceptibility}

\subsection{The Calculation Framework}

A method for calculating this quantity of R-current in the dual gravitational
description was formulated in Ref \cite{Son:2002sd}. In this paper,
we consider a spin system in quasi-2 dimension, a proper framework
for such system is not 2+1 but a 3+1 dimensional system in Minkovski
spacetime ($M_{4}$) as a direct consequence of the transverse nature
of magnetic fields, which will be shown in the following discussion
in more detail.

The large-N and large 't Hooft coupling limit of the theory in $M_{4}$
at finite temperature $T$ corresponds to a gravitational background
with a 4+1-dimensional asymptotically AdS metric embedding a Schwarzschild
blackhole\begin{equation}
ds^{2}=\frac{(\pi TR)^{2}}{u}\left[-f(u)dt^{2}+dx^{2}+dy^{2}+dz^{2}\right]+\frac{R^{2}}{4u^{2}f(u)}du^{2},\end{equation}
where $f(u)=1-u^{2}$, $u$ values from 0 (boundary) to 1 (horizon
with Hawkin temperature $T$), and R is the curvature radius of the
AdS space. According to the duality prescription, the two-point function
of spins in the holographic spin system is calculated by analyzing
linear corresponse of magnetic fields propagating in the 4+1-dimensional
AdS-Schwarzschild gravitational background. The perturbations of the
magnetic fields obey the Maxwell's equation\begin{equation}
\partial_{I}\left(\sqrt{-g}g^{IJ}g^{KL}F_{JL}\right)=0,\end{equation}
where $g_{IJ}$ is the metric of the background. The magnetic fields
defined as $B_{i}=\frac{1}{2}\epsilon_{ijk}F_{jk}$ is a pseudo-vector
propagating in the bulk space. The Fourier transformation of it by
its boundary coordinates is given by\begin{equation}
B_{i}(u,t,z)=\int\frac{d\omega dq_{z}}{(2\pi)^{2}}e^{iq_{z}z-i\omega t}B_{i}(u,\omega,q_{z}),\end{equation}
in which we have lied the wave vector $\mathbf{q}$ along the z-direction
for simplicity. By using the Bianchi identity for $F_{IJ}$ to relate
magnetic $\mathbf{B}$ fields to the electric $\mathbf{E}$ fields,
we obtain the wave equation\begin{equation}
B_{T}^{\prime\prime}+\frac{f^{\prime}}{f}B_{T}^{\prime}+\frac{\tilde{\omega}^{2}-\tilde{q}^{2}f}{uf^{2}}B_{T}=0,\label{eq:wave_equ}\end{equation}
in which $B_{T}=B_{x},B_{y}$ denotes the transverse components of
the magnetic fields, we also define dimensionless frequency and wave
vector \begin{equation}
\tilde{\omega}\equiv\frac{\omega}{2\pi T},\qquad\tilde{q}\equiv\frac{\left|q_{z}\right|}{2\pi T},\end{equation}
and the primes stand for derivatives with respect to $u$. 

Note that the magnetic fields are always perpendicular to the direction
of the wave vector (in the $z$-direction) due to their transverse
nature, so all the dual physical effects of the spin system are constrained
in the $x-y$ quasi-2-dimensional plane, which means the spin fluctuations
(duals to the transverse magnetic fields) are in fact constrained
in 2D layer, but it is {}``quasi-'' since it depends on external
parameters in the z dimension. For a fixed value of the external parameters,
the system can be viewed as pure 2D. Now let us turn to the Maxwell
action of the quasi-2D system in terms of magnetic fields\begin{align}
S_{cl}[\mathbf{B}] & =-\frac{N^{2}T^{2}}{16}V_{2}\int duf(u)\int\frac{d\omega dq_{z}}{(2\pi)^{2}}\frac{1}{q_{z}^{2}}\left[B_{x}^{\prime2}(u,\tilde{\omega},\tilde{q})+B_{y}^{\prime2}(u,\tilde{\omega},\tilde{q})+...\right]\end{align}
where $V_{2}$ is the area of this quasi-2D $x-y$ plane.

Applying the prescription proposed by Son and Starinets \cite{Son:2002sd},
using Eq.(\ref{eq:chi_from_S}), one finds the transverse spin susceptibility
of the isotropic quasi-2D system, ($\chi_{zz}=0$, $\chi_{xy}=\chi_{yx}=0$,
$\chi_{xx}=\chi_{yy}=\chi$)\begin{equation}
\chi=c\Phi(\tilde{\omega},\tilde{q}),\label{eq:chi_son}\end{equation}
in which $c$ is a dimensionless coefficient and the universal scaling
function $\Phi(\tilde{\omega},\tilde{q})$ governed by the conformal
nature of the holographic calculations \begin{align}
c & =\frac{N^{2}}{32\pi^{2}\tilde{q}^{2}},\qquad\Phi(\tilde{\omega},\tilde{q})=\lim_{u\rightarrow0}\frac{B_{T}^{\prime}(u,\tilde{\omega},\tilde{q})}{B_{T}(u,\tilde{\omega},\tilde{q})}.\label{eq:c_phi}\end{align}

Obviously, the $\chi$ is momentum-independent in the quasi-2D system
($\tilde{q}$ can be viewed as an external parameter or integration
constant determined from initial condition). It is very natural that
$\tilde{q}$ takes certain fixed value when $\tilde{\omega}$ is non-vanished
in ac susceptibility, unless uniform and static limit is required
to carefully take in dc susceptibility. So in the following discussions,
we are only interested in the frequency over temperature dependent,
i.e. $\Phi(\tilde{\omega})$.

One of the main purpose of the paper is to describe the critical scaling
behavior of the universal function $\Phi(\tilde{\omega})$ and/or
spin susceptibility, particularly in three exactly tractable asymptotics:
(i) The low frequency/high temperature hydrodynamic limit ($\tilde{\omega}\ll1$).
(ii) The high frequency/low temperature limit in which we have assumed
that the modes at high frequency are highly relativistic with linear
dispersion ($1\ll\tilde{\omega}=\tilde{q}$). (iii) The uniform and
static limit dc susceptibility ($\tilde{q},\tilde{\omega}\rightarrow0$).

\subsection{Low Frequency Hydrodynamic Limit ($\omega\ll T$)}

The low frequency/high temperature asymptotics is a straightforward
application of the perturbation theory to the wave equations Eq.(\ref{eq:wave_equ}).
The solution obeying the incoming wave boundary condition at the horizon
($u=1$) is controlled by a singular prefactor $(1-u)^{-i\tilde{\omega}/2}$.
Then the solution can be given perturbatively by using $\tilde{\omega}$
and $\tilde{q}$ as small expansion parameters\begin{equation}
B_{T}(u)=C(1-u)^{-i\tilde{\omega}/2}\left[1+\frac{i\tilde{\omega}}{2}\ln\frac{1+u}{2}+\frac{\tilde{q}^{2}}{2}\left(\frac{\pi^{2}}{2}+\mathrm{Li_{2}}(-u)+\ln u\ln(1+u)+\mathrm{Li_{2}}(1-u)\right)+\mathcal{O}(\tilde{\omega}^{2},\tilde{q}^{4},\tilde{\omega}\tilde{q}^{2})\right],\end{equation}
where the renormalization constant $C(\tilde{\omega},\tilde{q})$
is determined by the boundary condition $\lim_{u\rightarrow0}B_{T}(u)=B_{T}^{0}$,\begin{equation}
C(\tilde{\omega},\tilde{q})=\frac{8B_{T}^{0}}{8-4i\tilde{\omega}\ln2+\pi^{2}\tilde{q}^{2}}.\end{equation}
we get the derivative of the B fields on the boundary ($u\rightarrow0$),
\begin{equation}
\lim_{u\rightarrow0}B_{T}^{\prime}=i\tilde{\omega}B_{T}^{0}.\end{equation}
At lowest order, the result is momentum independent, so the spin susceptibility
has a simple scaling behavior $\Phi_{h}(\tilde{\omega})=i\tilde{\omega}$,
this can be seem from Eq.(\ref{eq:chi_son}) and Eq.(\ref{eq:c_phi}),
\begin{equation}
\chi_{h}(\tilde{\omega})=ci\tilde{\omega},\qquad(\tilde{\omega}\ll1).\label{eq:chi_hydro}\end{equation}

We will see in the next section that this $\alpha=1$ dynamic scaling
law at low frequency/high temperature limit is very important in understanding
the behaviors of a quantum critical regime, which is considered as
a finite temperature extension of a quantum critical point.

\subsection{High Frequency Relativistic Limit ($T\ll\omega=q$)}

To investigate the quantum critical point at near zero temperature,
we need to study the high frequency/low temperature limit. This limit
of the solution requires a careful WKB analysis. In this subsection,
we assume that in this high frequency limit, the modes are relativistic
with linear dispersion $\omega=q$, so the Eq.(\ref{eq:wave_equ})
becomes \begin{equation}
B_{T}^{\prime\prime}+\frac{f^{\prime}}{f}B_{T}^{\prime}+\frac{u\tilde{\omega}^{2}}{f^{2}}B_{T}=0.\label{eq:wave_equ_rela}\end{equation}
For $\tilde{\omega}\gg1$, by using Langer-Olver's method \cite{1954RSPTA.247..307O,Policastro:2001yb},
we are able to obtain uniform asymptotics expansions to the solution
(a version of the WKB approximation). If we introduce a new variable,\begin{equation}
B_{T}=\frac{1}{\sqrt{-f(u)}}\phi,\end{equation}
then the wave equation is rewritten as\begin{equation}
\phi^{\prime\prime}=-\frac{u\tilde{\omega}^{2}+1}{(1-u^{2})^{2}}\phi.\end{equation}
For large values of $\tilde{\omega}$, the solution has the formal
expansions in terms of Airy function, which is chosen to obey the
incoming wave boundary condition,\begin{equation}
B_{T}(u)\sim\frac{C}{\sqrt{-f(u)}}\left[\frac{-u}{(1-u^{2})^{2}\zeta(-u)}\right]^{-1/4}\mathrm{Ai}\left(\tilde{\omega}^{2/3}\zeta(-u)\right)+...\end{equation}
where $\mathrm{Ai}(z)$ is the Airy function, and\begin{equation}
\zeta(x)=\frac{3^{2/3}}{2^{4/3}}\left(i\pi-2\arctan\sqrt{x}+\ln\frac{\sqrt{x}+1}{\sqrt{x}-1}\right).\end{equation}
The renormalization constant $C(\tilde{\omega})$ is\begin{equation}
C(\tilde{\omega})=2\sqrt{\pi}e^{i\pi/4}\tilde{\omega}^{1/6}2^{-i\tilde{\omega}-\tilde{\omega}/2}e^{i\pi\tilde{\omega}/4}B_{T}^{0}.\end{equation}
Similarly, by using Eq.(\ref{eq:c_phi}), a fractional exponent $\alpha=2/3$
is finally obtained, i.e. $\Phi_{r}(\tilde{\omega})\propto\tilde{\omega}^{2/3}$,
and\begin{equation}
\chi_{r}(\tilde{\omega})\sim c\frac{3^{1/3}\Gamma\left(\frac{2}{3}\right)}{\Gamma\left(\frac{1}{3}\right)}\left(-\tilde{\omega}\right)^{2/3},\qquad(\tilde{\omega}\gg1).\label{eq:chi_relati}\end{equation}

Different from the low frequency limit where the scaling is momentum-independence,
in fact, numerical study manifests that the scaling function $\Phi(\tilde{\omega},\tilde{q})$
is sensitive to $\tilde{q}$ in the high frequency limit. At first
sight the $\alpha=2/3$ scaling we obtained here is achieved by taking
fine-tuned linear dispersion in Eq.(\ref{eq:wave_equ_rela}), but
it shows that the dispersion of vector mode at high frequency tends
to linear \cite{PhysRevD.67.124013}, so $\omega=q$ at high frequency
limit is expected as a promising pre-assumption for most real cases.

\subsection{DC Susceptibility or Uniform Static Limit ($q,\omega\rightarrow0$)}

In principle the uniform ($q\rightarrow0$) and static ($\omega\rightarrow0$)
limit can be taken straightforwardly in the low frequency hydrodynamic
regime,\begin{align}
\chi_{s} & =\lim_{\omega,q\rightarrow0}\chi_{h}(\tilde{\omega},\tilde{q}).\end{align}
However, the $\tilde{q}$ dependence of the prefactor $c$ requires
a more careful treatment, since $\tilde{q}$ of the modes will also
tend to vanish as $\tilde{\omega}$ goes to zero because of the dispersion
relation. The limit exists and be finite, if we note that $\tilde{\omega}$
and $\tilde{q}^{2}$ are of the same order in this hydrodynamic regime,
and more precisely, there is a diffusive pole $i\tilde{\omega}=\tilde{q}^{2}$
($i\omega=Dq^{2}$ with $D=1/2\pi T$) governs the low energy dispersion
for the longitudinal vector modes who share the same wave vector and
frequency with the transverse vector modes in hydrodynamic regime.
Therefore, we obtain a universal real constant susceptibility \cite{2011JHEP...11..142L},
independent with temperature,\begin{equation}
\chi_{s}=\frac{N^{2}}{32\pi^{2}}\lim_{\omega,q\rightarrow0}\frac{i\tilde{\omega}}{\tilde{q}^{2}}=\frac{N^{2}}{32\pi^{2}}.\label{eq:chi_static}\end{equation}

The non-zero dimensionless constant dc spin susceptibility at high
temperature for strongly coupled critical spin fluctuations is a non-trivial
prediction from holographic theory at the critical regime. The universality
of the strongly coupled critical spin susceptibility measuring the
critical spin transport is a direct consequence of the {}``perfect
fluid'' behavior of a critical matter, which may be analogous to
the universal value of the shear viscosity over entropy density of
the strongly coupled quark-gluon plasma around critical temperature
observed in heavy-ion-collision.

\section{\label{sec:Connection-to-experiments}Connection to experiments}

The spin susceptibility we obtained from holographic large-N calculation
is sensitive to the total number of degrees of freedom N, and hence
it is not a good quantity to compare with measurements, but in this
paper we propose two ways to connect our calculations to experimental
facts: (i) the scaling exponent is universal and N-independent at
lowest order in the holographic calculations, so it can be directly
compared to experiments. (ii) The itinerant electrons coupling to
the spin fluctuations give rise to a contributions $1/N^{2}$ in large-N
theory, recalling that the order of spin susceptibility Eq.(\ref{eq:c_phi})
is $N^{2}$, this reflects that the resistivity and self-energy of
the N-component electrons due to exchange the spin fluctuations are
of order $N^{0}$, and hence they are manifested in insensitive to
the large-N technique.

\subsection{Scaling of Critical Spin Susceptibility in Experiments}

The results of scaling behavior for strongly coupled critical spin
fluctuations lead to a variety of direct measurement of critical spin
susceptibility by neutron-scattering, NMR, and magnetometry measurements.
The low frequency hydrodynamic limit of the system can be measured
in the phase diagram at large range of temperatures close to the doping/fields
induced criticality, i.e. the QCR. The Eq.(\ref{eq:chi_hydro}) suggests
that, when frequency of the ac magnetic fields is fixed, the spin
susceptibility decreases as the temperature increases. The simple
$\omega/T$ scaling of spin susceptibility at low frequency and/or
high temperature proposed in Eq.(\ref{eq:chi_hydro}) is observed
in $\mathrm{La_{2-x}Sr_{x}CuO_{4}}$ compound (at critical doping
$x\approx0.04\pm0.01$) in the pseudogap regime (which is conjectured
as a QCR) \cite{PhysRevLett.67.1930}, and it is also suggested as
a phenomenological description in normal state of Cu-O high-Tc superconductor
\cite{PhysRevLett.63.1996}, \begin{equation}
\mathrm{Im}\chi\sim\frac{\omega}{T},\qquad(\omega\ll T).\end{equation}

When temperature is driven to small value, at the critical doping
or QCP, the fluctuations become strong and almost quantum critical.
In this case, we are interested in the $\chi\sim T^{-\alpha}$ type
of scaling measured at fixed frequency, which is considered associated
with the proximity to a QCP. We find the spin susceptibility diverges
governed by a fractional scaling exponent $\alpha=2/3$ in the high
frequency or low temperature regime. This prediction of scaling from
Eq.(\ref{eq:chi_relati}) is supported from the experimental observation
in e.g. $\mathrm{YbRh_{2}(Si_{1-x}Ge_{x})_{2}}$ ($x\approx0.05$)
when reached a QCP. Although $\chi(T)$ tends to saturation below
Kondo temperature, approaching to QCP from $0.3K$ to $10K$, it can
be approximated by a power-law divergence \cite{PhysRevLett.94.076402},
\begin{equation}
\left(\chi(T)-\mathrm{const}\right)\sim T^{-0.6},\qquad(T\sim0.3K\div10K),\end{equation}
where $\mathrm{const}=0.215\times10^{-6}\mathrm{m^{3}mol^{-1}}$ is
a small temperature-independent contribution. We can see at high $T$
, $\chi\rightarrow\mathrm{const}$. In our framework the constant
may be interpreted as a constant lower bound of static spin susceptibility
as suggested in Eq.(\ref{eq:chi_static}). 

The fractional scaling law of temperature has also been seen in the
kagome lattice anti-ferromagnetic herbertsmithite $\mathrm{ZnCu_{3}(OH)_{6}Cl_{2}}$
at fixed frequency of weak ac magnetic fields, which is thought of
possibly displaying quantum critical behavior and the proximity to
a critical spin liquid ground state. For both real and imaginary part
of susceptibility, from $0K$ to $35K$, the divergence scaling behaves
as \cite{PhysRevLett.104.147201} \begin{equation}
\chi(T)\sim T^{-(0.66\pm0.02)},\qquad(T\sim0K\div35K).\end{equation}

Other neutron scattering experiments on $\mathrm{CeCu_{6-x}Au_{x}}$
near the critical doping $x_{c}\approx0.1$ also gives an simple $\omega/T$
scaling and a similar fractional critical exponent, fitting $\chi\sim T^{-\alpha}$
with $\alpha\approx0.75$ closed to our result \cite{PhysRevLett.75.725}. 

Within the experimental resolution, it is a good agreement in our
calculation to these measurements. Note that such fractional scaling
can not be explained by any perturbative spin-density-wave QCP theories,
and hence be a good support for our holographic treatment of the critical
spin fluctuations. The author also recognize that there are other
candidates, e.g. non-perturbative phenomenological approach \cite{PhysRevLett.94.066402}
and dynamical mean field theory \cite{2001Natur.413..804S} yielding
the similar fractional exponent.

\subsection{Linear-T Resistivity in Normal State of Cuprates}

The normal state or strange metal phase of the cuprate is regarded
as the central dogma in the theory of high-Tc superconductivity. Different
from the $R\sim T^{2}$ behavior of resistivity due to phonon scattering
in usual Fermi liquid metals, in cuprate layers, a linear-T resistivity
$R\sim T$ near optimal doping (as critical doping) over a wide range
of temperature is a generic property of the normal state. The fact
that the constant of proportionality seems similar for different cuprates
may be a hint suggesting such behavior is closely related to the universality
of critical phenomenon. We will show that the scaling behavior of
spin susceptibility at low frequency/high temperature limit is very
important in understanding the anomalous behavior, if we replace the
phonons by the critical spin fluctuations. The $\alpha=1$ scaling
of $\mathrm{Im}\chi$ at low frequency is physically similar with
the phenomenological Marginal-Fermi-Liquid (MFL) theory \cite{PhysRevLett.63.1996}
in two aspects, (i) momentum-independence of the spin susceptibility
at generic wave vector in the quasi-2D spin system, (ii) scaling invariant
form of the spin susceptibility due to the CFT nature of the holographic
spin system.

Here we consider Kondo lattice model, in which the coupling between
the conducting electrons to the critical spin fluctuations is given
by the Kondo exchange interaction in large-N,\begin{equation}
H_{I}=\frac{J_{K}}{N}\sum_{k}\psi_{k}^{\dagger}\left(\psi_{k}\mathbf{\sigma}\cdot\mathbf{S}_{k}\right),\end{equation}
where $J_{K}/N$ is the coupling constant, $\psi$ is the N-component
electron, $\mathbf{\sigma}$ is the Pauli matrix and $\mathbf{S}$
is the spin operator of the spin fluctuation. We treat this interaction
perturbatively which is reasonable in high-T strange metal phase due
to Kondo screening. The self-energy of electron due to exchange the
critical spin fluctuations is then given by\begin{equation}
\Sigma(\epsilon,k)=\frac{J_{K}^{2}}{N^{2}}\sum_{q}\int\frac{d\omega}{2\pi}\int\frac{d\omega^{\prime}}{2\pi}\frac{A(q,\omega)B(k-q,\omega^{\prime})}{\epsilon-\omega-\omega^{\prime}}\left[n(\omega^{\prime})+1-f(\omega)\right],\end{equation}
in which $A=-2\mathrm{Im}G_{0}^{R}$ is the spectral function of free
electron, $B=-2\mathrm{Im}\chi$ is the spectral function of the critical
spin fluctuations, $n,f$ are Bose (for spin) and Fermi (for electron)
distribution functions, respectively. For a $\mathbf{q}$-independent
spectrum given in the spin susceptibility at low frequency, the momentum
integration can be performed directly to $A(q,\omega)$, which gives
the density of states for electrons $\rho(\omega)=\frac{1}{2\pi V}\sum_{q}A(q,\omega)$.
Then we use the Kramers-Kronig relations to dealing with the integration
of $\omega$, we get the imaginary part of $\Sigma$,\begin{equation}
\mathrm{Im}\Sigma(\epsilon)=\frac{J_{K}^{2}}{N^{2}}\rho_{F}\pi\int\frac{d\omega^{\prime}}{2\pi}B(\omega^{\prime})\left[n(\omega^{\prime})+f(\omega^{\prime}-\epsilon)\right],\end{equation}
in which the density of state is approximately a constant density
in a unit volume at the Fermi surface $\rho_{F}=\rho(0)$. We extend
the validity of $B=-2\mathrm{Im}\chi=-2c\tilde{\omega}$ to all range
of frequency in our integral, since here it is treated in the high-T
strange metal phase. A straightforward calculation then gives\begin{equation}
\mathrm{Im}\Sigma(\epsilon,T)=\frac{J_{K}^{2}}{N^{2}}\rho_{F}cT\mathcal{F}(\frac{\epsilon}{T}),\label{eq:im_self}\end{equation}
where $\mathcal{F}(\epsilon/T)$ is given by\begin{equation}
\mathcal{F}(\frac{\epsilon}{T})=\frac{\pi^{2}}{4}+\frac{\epsilon}{T}\ln2+...\qquad(\epsilon\ll T)\end{equation}
which approaches a constant for small $\epsilon/T$, we have\begin{equation}
\mathrm{Im}\Sigma(\epsilon,T)=\lambda T,\qquad(\epsilon\ll T)\end{equation}
where $\lambda=\frac{1}{128}J_{K}^{2}\rho_{F}/\tilde{q}^{2}$ is a
dimensionless effective coupling constant of order $N^{0}$. By using
the Kramers-Kronig relations again we obtain the real part and the
full self-energy from the imaginary part\begin{equation}
\Sigma(\epsilon,T)=-\frac{2}{\pi}\lambda\left(\epsilon\ln\left|\frac{x}{\epsilon_{c}}\right|-i\frac{\pi}{2}x\right),\end{equation}
where $x=\mathrm{max}(\epsilon,T)$, $\epsilon_{c}$ is a ultraviolet
cutoff scale. This form of self-energy has been proposed in the Marginal-Fermi-Liquid
\cite{PhysRevLett.63.1996} and fits well with the ARPES measurement
of cuprates in its high-T strange metal phase \cite{2000PNAS...97.5714A}.

According to the Optical Theorem, the imaginary part of the self-energy
is related to the scattering amplitude between the free electron and
the critical spin fluctuations. Therefore, in the quasi-2D plane,
we find the electric dc resistivity behaves linear in temperature,\begin{equation}
R=\lambda T.\end{equation}

It is a generic property observed in normal state, strange metal and
many non-Fermi liquids materials in 2D conductive layer, which are
expected to exhibit QCR behavior related to the optimal doping, rather
a $T^{2}$ behavior in Fermi liquids. In this sense, we could conclude
that the non-Fermi liquids could arise from coupling a Fermi liquids
to a strongly coupled critical spin fluctuations.

\section{\label{sec:Conclusions}Conclusions}

We conclude this paper by recollecting some highlights to the problem
of the strongly coupled critical spin fluctuations. The dynamics of
spin fluctuations is very important in understanding the behavior
of normal state of the heavy fermion metal and high-Tc superconductor.
The strongly coupled spin fluctuations develop quantum criticality
and the notion of quasi-particle is no longer valid. Such critical
system is described by continuum field theory in the vicinity of a
non-Gaussian fixed point with strong coupling. The treatment of the
system is beyond the traditional perturbative technique. We present
a holographic theory to the system by following the AdS/CFT correspondence
which conjectures that the strongly coupled critical system is dual
to a gravitational theory in AdS space at large-N limit. We calculate
the spin susceptibility in such holographic system at large 't Hooft
coupling, and find that (i) the holographic spin system is quasi-2-dimensional;
(ii) the $\omega/T$ scaling of spin susceptibility independent to
wave vector is a general property of holographic spin fluctuations
in the quasi-2-dimensional system; (iii) the scaling behavior of ac
spin susceptibility is universal, $\chi\sim\omega/T$ at low frequency
hydrodynamic limit and $\chi\sim(\omega/T)^{2/3}$ at high frequency
relativistic limit, which can not be explained by traditional perturbative
spin-density-wave QCP theories; (iv) the dc spin susceptibility approaches
a constant $N^{2}/32\pi^{2}$ independent to temperature; (v) these
scaling law shown in the susceptibility agree well with a number of
experimental measurements, in which the test materials are tuned to
nearly critical doping where quantum critical point/quantum critical
regime appears; (vi) The $\mathrm{Im}\chi\sim\omega/T$ scaling at
low frequency/high temperature limit gives rise to a linear-temperature
resistivity in strange metals and/or normal state of high-Tc superconductor.
We argue that the non-Fermi liquid can arise from coupling the Fermi
liquid to strongly coupled critical spin fluctuations.
\begin{acknowledgments}
The author would like to thank S.Sachdev, S.Hartnoll and H.Liu for
helpful communications.
\end{acknowledgments}
\bibliographystyle{apsrev}

\begin{thebibliography}{53}
\expandafter\ifx\csname natexlab\endcsname\relax\def\natexlab#1{#1}\fi
\expandafter\ifx\csname bibnamefont\endcsname\relax
  \def\bibnamefont#1{#1}\fi
\expandafter\ifx\csname bibfnamefont\endcsname\relax
  \def\bibfnamefont#1{#1}\fi
\expandafter\ifx\csname citenamefont\endcsname\relax
  \def\citenamefont#1{#1}\fi
\expandafter\ifx\csname url\endcsname\relax
  \def\url#1{\texttt{#1}}\fi
\expandafter\ifx\csname urlprefix\endcsname\relax\def\urlprefix{URL }\fi
\providecommand{\bibinfo}[2]{#2}
\providecommand{\eprint}[2][]{\url{#2}}

\bibitem[{\citenamefont{Manousakis}(1991)}]{RevModPhys.63.1}
\bibinfo{author}{\bibfnamefont{E.}~\bibnamefont{Manousakis}},
  \bibinfo{journal}{Rev. Mod. Phys.} \textbf{\bibinfo{volume}{63}},
  \bibinfo{pages}{1} (\bibinfo{year}{1991}).

\bibitem[{\citenamefont{{Balents}}(2010)}]{2010Natur.464..199B}
\bibinfo{author}{\bibfnamefont{L.}~\bibnamefont{{Balents}}},
  \bibinfo{journal}{\nat} \textbf{\bibinfo{volume}{464}}, \bibinfo{pages}{199}
  (\bibinfo{year}{2010}).

\bibitem[{\citenamefont{{Coleman}}(2006)}]{2006cond.mat.12006C}
\bibinfo{author}{\bibfnamefont{P.}~\bibnamefont{{Coleman}}},
  \bibinfo{journal}{eprint arXiv:cond-mat/0612006}  (\bibinfo{year}{2006}),
  \eprint{arXiv:cond-mat/0612006}.

\bibitem[{\citenamefont{Scalapino et~al.}(1986)\citenamefont{Scalapino, Loh,
  and Hirsch}}]{PhysRevB.34.8190}
\bibinfo{author}{\bibfnamefont{D.~J.} \bibnamefont{Scalapino}},
  \bibinfo{author}{\bibfnamefont{E.}~\bibnamefont{Loh}}, \bibnamefont{and}
  \bibinfo{author}{\bibfnamefont{J.~E.} \bibnamefont{Hirsch}},
  \bibinfo{journal}{Phys. Rev. B} \textbf{\bibinfo{volume}{34}},
  \bibinfo{pages}{8190} (\bibinfo{year}{1986}).

\bibitem[{\citenamefont{{Gegenwart} et~al.}(2008)\citenamefont{{Gegenwart},
  {Si}, and {Steglich}}}]{2008NatPh...4..186G}
\bibinfo{author}{\bibfnamefont{P.}~\bibnamefont{{Gegenwart}}},
  \bibinfo{author}{\bibfnamefont{Q.}~\bibnamefont{{Si}}}, \bibnamefont{and}
  \bibinfo{author}{\bibfnamefont{F.}~\bibnamefont{{Steglich}}},
  \bibinfo{journal}{Nature Physics} \textbf{\bibinfo{volume}{4}},
  \bibinfo{pages}{186} (\bibinfo{year}{2008}), \eprint{0712.2045}.

\bibitem[{\citenamefont{{Schr{\"o}der}
  et~al.}(2000)\citenamefont{{Schr{\"o}der}, {Aeppli}, {Coldea}, {Adams},
  {Stockert}, {L{\"o}hneysen}, {Bucher}, {Ramazashvili}, and
  {Coleman}}}]{2000Natur.407..351S}
\bibinfo{author}{\bibfnamefont{A.}~\bibnamefont{{Schr{\"o}der}}},
  \bibinfo{author}{\bibfnamefont{G.}~\bibnamefont{{Aeppli}}},
  \bibinfo{author}{\bibfnamefont{R.}~\bibnamefont{{Coldea}}},
  \bibinfo{author}{\bibfnamefont{M.}~\bibnamefont{{Adams}}},
  \bibinfo{author}{\bibfnamefont{O.}~\bibnamefont{{Stockert}}},
  \bibinfo{author}{\bibfnamefont{H.~v.} \bibnamefont{{L{\"o}hneysen}}},
  \bibinfo{author}{\bibfnamefont{E.}~\bibnamefont{{Bucher}}},
  \bibinfo{author}{\bibfnamefont{R.}~\bibnamefont{{Ramazashvili}}},
  \bibnamefont{and}
  \bibinfo{author}{\bibfnamefont{P.}~\bibnamefont{{Coleman}}},
  \bibinfo{journal}{\nat} \textbf{\bibinfo{volume}{407}}, \bibinfo{pages}{351}
  (\bibinfo{year}{2000}), \eprint{arXiv:cond-mat/0011002}.

\bibitem[{\citenamefont{{Sachdev}}(2008)}]{2008NatPh...4..173S}
\bibinfo{author}{\bibfnamefont{S.}~\bibnamefont{{Sachdev}}},
  \bibinfo{journal}{Nature Physics} \textbf{\bibinfo{volume}{4}},
  \bibinfo{pages}{173} (\bibinfo{year}{2008}), \eprint{0711.3015}.

\bibitem[{\citenamefont{Chubukov et~al.}(1994)\citenamefont{Chubukov, Sachdev,
  and Ye}}]{PhysRevB.49.11919}
\bibinfo{author}{\bibfnamefont{A.~V.} \bibnamefont{Chubukov}},
  \bibinfo{author}{\bibfnamefont{S.}~\bibnamefont{Sachdev}}, \bibnamefont{and}
  \bibinfo{author}{\bibfnamefont{J.}~\bibnamefont{Ye}}, \bibinfo{journal}{Phys.
  Rev. B} \textbf{\bibinfo{volume}{49}}, \bibinfo{pages}{11919}
  (\bibinfo{year}{1994}).

\bibitem[{\citenamefont{{Shaginyan} et~al.}(2010)\citenamefont{{Shaginyan},
  {Amusia}, {Msezane}, and {Popov}}}]{2010PhR...492...31S}
\bibinfo{author}{\bibfnamefont{V.~R.} \bibnamefont{{Shaginyan}}},
  \bibinfo{author}{\bibfnamefont{M.~Y.} \bibnamefont{{Amusia}}},
  \bibinfo{author}{\bibfnamefont{A.~Z.} \bibnamefont{{Msezane}}},
  \bibnamefont{and} \bibinfo{author}{\bibfnamefont{K.~G.}
  \bibnamefont{{Popov}}}, \bibinfo{journal}{Phys.Rep.}
  \textbf{\bibinfo{volume}{492}}, \bibinfo{pages}{31} (\bibinfo{year}{2010}),
  \eprint{arXiv:1006.2658}.

\bibitem[{\citenamefont{{Laughlin} et~al.}(2001)\citenamefont{{Laughlin},
  {Lonzarich}, {Monthoux}, and {Pines}}}]{2001AdPhy..50..361L}
\bibinfo{author}{\bibfnamefont{R.~B.} \bibnamefont{{Laughlin}}},
  \bibinfo{author}{\bibfnamefont{G.~G.} \bibnamefont{{Lonzarich}}},
  \bibinfo{author}{\bibfnamefont{P.}~\bibnamefont{{Monthoux}}},
  \bibnamefont{and} \bibinfo{author}{\bibfnamefont{D.}~\bibnamefont{{Pines}}},
  \bibinfo{journal}{Advances in Physics} \textbf{\bibinfo{volume}{50}},
  \bibinfo{pages}{361} (\bibinfo{year}{2001}).

\bibitem[{\citenamefont{{Custers} et~al.}(2003)\citenamefont{{Custers},
  {Gegenwart}, {Wilhelm}, {Neumaier}, {Tokiwa}, {Trovarelli}, {Geibel},
  {Steglich}, {P{\'e}pin}, and {Coleman}}}]{2003Natur.424..524C}
\bibinfo{author}{\bibfnamefont{J.}~\bibnamefont{{Custers}}},
  \bibinfo{author}{\bibfnamefont{P.}~\bibnamefont{{Gegenwart}}},
  \bibinfo{author}{\bibfnamefont{H.}~\bibnamefont{{Wilhelm}}},
  \bibinfo{author}{\bibfnamefont{K.}~\bibnamefont{{Neumaier}}},
  \bibinfo{author}{\bibfnamefont{Y.}~\bibnamefont{{Tokiwa}}},
  \bibinfo{author}{\bibfnamefont{O.}~\bibnamefont{{Trovarelli}}},
  \bibinfo{author}{\bibfnamefont{C.}~\bibnamefont{{Geibel}}},
  \bibinfo{author}{\bibfnamefont{F.}~\bibnamefont{{Steglich}}},
  \bibinfo{author}{\bibfnamefont{C.}~\bibnamefont{{P{\'e}pin}}},
  \bibnamefont{and}
  \bibinfo{author}{\bibfnamefont{P.}~\bibnamefont{{Coleman}}},
  \bibinfo{journal}{\nat} \textbf{\bibinfo{volume}{424}}, \bibinfo{pages}{524}
  (\bibinfo{year}{2003}), \eprint{arXiv:cond-mat/0308001}.

\bibitem[{\citenamefont{{Sachdev} and {Keimer}}(2011)}]{2011PhT....64b..29S}
\bibinfo{author}{\bibfnamefont{S.}~\bibnamefont{{Sachdev}}} \bibnamefont{and}
  \bibinfo{author}{\bibfnamefont{B.}~\bibnamefont{{Keimer}}},
  \bibinfo{journal}{Physics Today} \textbf{\bibinfo{volume}{64}},
  \bibinfo{pages}{020000} (\bibinfo{year}{2011}), \eprint{1102.4628}.

\bibitem[{\citenamefont{Maldacena}(1998)}]{Maldacena:1997re}
\bibinfo{author}{\bibfnamefont{J.~M.} \bibnamefont{Maldacena}},
  \bibinfo{journal}{Adv. Theor. Math. Phys.} \textbf{\bibinfo{volume}{2}},
  \bibinfo{pages}{231} (\bibinfo{year}{1998}), \eprint{hep-th/9711200}.

\bibitem[{\citenamefont{Witten}(1998)}]{Witten:1998qj}
\bibinfo{author}{\bibfnamefont{E.}~\bibnamefont{Witten}},
  \bibinfo{journal}{Adv. Theor. Math. Phys.} \textbf{\bibinfo{volume}{2}},
  \bibinfo{pages}{253} (\bibinfo{year}{1998}), \eprint{hep-th/9802150}.

\bibitem[{\citenamefont{Policastro et~al.}(2001)\citenamefont{Policastro, Son,
  and Starinets}}]{Policastro:2001yc}
\bibinfo{author}{\bibfnamefont{G.}~\bibnamefont{Policastro}},
  \bibinfo{author}{\bibfnamefont{D.~T.} \bibnamefont{Son}}, \bibnamefont{and}
  \bibinfo{author}{\bibfnamefont{A.~O.} \bibnamefont{Starinets}},
  \bibinfo{journal}{Phys. Rev. Lett.} \textbf{\bibinfo{volume}{87}},
  \bibinfo{pages}{081601} (\bibinfo{year}{2001}), \eprint{hep-th/0104066}.

\bibitem[{\citenamefont{Policastro et~al.}(2002)\citenamefont{Policastro, Son,
  and Starinets}}]{Policastro:2002se}
\bibinfo{author}{\bibfnamefont{G.}~\bibnamefont{Policastro}},
  \bibinfo{author}{\bibfnamefont{D.~T.} \bibnamefont{Son}}, \bibnamefont{and}
  \bibinfo{author}{\bibfnamefont{A.~O.} \bibnamefont{Starinets}},
  \bibinfo{journal}{JHEP} \textbf{\bibinfo{volume}{09}}, \bibinfo{pages}{043}
  (\bibinfo{year}{2002}), \eprint{hep-th/0205052}.

\bibitem[{\citenamefont{Gavin and Abdel-Aziz}(2006)}]{PhysRevLett.97.162302}
\bibinfo{author}{\bibfnamefont{S.}~\bibnamefont{Gavin}} \bibnamefont{and}
  \bibinfo{author}{\bibfnamefont{M.}~\bibnamefont{Abdel-Aziz}},
  \bibinfo{journal}{Phys. Rev. Lett.} \textbf{\bibinfo{volume}{97}},
  \bibinfo{pages}{162302} (\bibinfo{year}{2006}).

\bibitem[{\citenamefont{{Hartnoll}}(2009)}]{2009CQGra..26v4002H}
\bibinfo{author}{\bibfnamefont{S.~A.} \bibnamefont{{Hartnoll}}},
  \bibinfo{journal}{Classical and Quantum Gravity}
  \textbf{\bibinfo{volume}{26}}, \bibinfo{pages}{224002}
  (\bibinfo{year}{2009}), \eprint{0903.3246}.

\bibitem[{\citenamefont{{Sachdev}}(2010)}]{2010arXiv1002.2947S}
\bibinfo{author}{\bibfnamefont{S.}~\bibnamefont{{Sachdev}}},
  \bibinfo{journal}{ArXiv e-prints}  (\bibinfo{year}{2010}),
  \eprint{1002.2947}.

\bibitem[{\citenamefont{Hartnoll
  et~al.}(2008{\natexlab{a}})\citenamefont{Hartnoll, Herzog, and
  Horowitz}}]{Hartnoll:2008vx}
\bibinfo{author}{\bibfnamefont{S.~A.} \bibnamefont{Hartnoll}},
  \bibinfo{author}{\bibfnamefont{C.~P.} \bibnamefont{Herzog}},
  \bibnamefont{and} \bibinfo{author}{\bibfnamefont{G.~T.}
  \bibnamefont{Horowitz}}, \bibinfo{journal}{Phys.Rev.Lett.}
  \textbf{\bibinfo{volume}{101}}, \bibinfo{pages}{031601}
  (\bibinfo{year}{2008}{\natexlab{a}}), \eprint{0803.3295}.

\bibitem[{\citenamefont{Hartnoll
  et~al.}(2008{\natexlab{b}})\citenamefont{Hartnoll, Herzog, and
  Horowitz}}]{Hartnoll:2008kx}
\bibinfo{author}{\bibfnamefont{S.~A.} \bibnamefont{Hartnoll}},
  \bibinfo{author}{\bibfnamefont{C.~P.} \bibnamefont{Herzog}},
  \bibnamefont{and} \bibinfo{author}{\bibfnamefont{G.~T.}
  \bibnamefont{Horowitz}}, \bibinfo{journal}{JHEP}
  \textbf{\bibinfo{volume}{0812}}, \bibinfo{pages}{015}
  (\bibinfo{year}{2008}{\natexlab{b}}), \eprint{0810.1563}.

\bibitem[{\citenamefont{Herzog et~al.}(2009)\citenamefont{Herzog, Kovtun, and
  Son}}]{Herzog:2008he}
\bibinfo{author}{\bibfnamefont{C.}~\bibnamefont{Herzog}},
  \bibinfo{author}{\bibfnamefont{P.}~\bibnamefont{Kovtun}}, \bibnamefont{and}
  \bibinfo{author}{\bibfnamefont{D.}~\bibnamefont{Son}},
  \bibinfo{journal}{Phys.Rev.} \textbf{\bibinfo{volume}{D79}},
  \bibinfo{pages}{066002} (\bibinfo{year}{2009}).

\bibitem[{\citenamefont{Herzog}(2009)}]{Herzog:2009xv}
\bibinfo{author}{\bibfnamefont{C.~P.} \bibnamefont{Herzog}},
  \bibinfo{journal}{J.Phys.A} \textbf{\bibinfo{volume}{A42}},
  \bibinfo{pages}{343001} (\bibinfo{year}{2009}).

\bibitem[{\citenamefont{Faulkner et~al.}(2010)\citenamefont{Faulkner, Iqbal,
  Liu, McGreevy, and Vegh}}]{Faulkner:2010da}
\bibinfo{author}{\bibfnamefont{T.}~\bibnamefont{Faulkner}},
  \bibinfo{author}{\bibfnamefont{N.}~\bibnamefont{Iqbal}},
  \bibinfo{author}{\bibfnamefont{H.}~\bibnamefont{Liu}},
  \bibinfo{author}{\bibfnamefont{J.}~\bibnamefont{McGreevy}}, \bibnamefont{and}
  \bibinfo{author}{\bibfnamefont{D.}~\bibnamefont{Vegh}}
  (\bibinfo{year}{2010}).

\bibitem[{\citenamefont{Sachdev}(2010{\natexlab{a}})}]{Sachdev:2010uj}
\bibinfo{author}{\bibfnamefont{S.}~\bibnamefont{Sachdev}},
  \bibinfo{journal}{J.Stat.Mech.} \textbf{\bibinfo{volume}{1011}},
  \bibinfo{pages}{P11022} (\bibinfo{year}{2010}{\natexlab{a}}),
  \eprint{1010.0682}.

\bibitem[{\citenamefont{{Faulkner} et~al.}(2010)\citenamefont{{Faulkner},
  {Iqbal}, {Liu}, {McGreevy}, and {Vegh}}}]{2010Sci...329.1043F}
\bibinfo{author}{\bibfnamefont{T.}~\bibnamefont{{Faulkner}}},
  \bibinfo{author}{\bibfnamefont{N.}~\bibnamefont{{Iqbal}}},
  \bibinfo{author}{\bibfnamefont{H.}~\bibnamefont{{Liu}}},
  \bibinfo{author}{\bibfnamefont{J.}~\bibnamefont{{McGreevy}}},
  \bibnamefont{and} \bibinfo{author}{\bibfnamefont{D.}~\bibnamefont{{Vegh}}},
  \bibinfo{journal}{Science} \textbf{\bibinfo{volume}{329}},
  \bibinfo{pages}{1043} (\bibinfo{year}{2010}).

\bibitem[{\citenamefont{Lee}(2009)}]{Lee:2008xf}
\bibinfo{author}{\bibfnamefont{S.-S.} \bibnamefont{Lee}},
  \bibinfo{journal}{Phys.Rev.} \textbf{\bibinfo{volume}{D79}},
  \bibinfo{pages}{086006} (\bibinfo{year}{2009}), \eprint{0809.3402}.

\bibitem[{\citenamefont{Liu et~al.}(2011)\citenamefont{Liu, McGreevy, and
  Vegh}}]{Liu:2009dm}
\bibinfo{author}{\bibfnamefont{H.}~\bibnamefont{Liu}},
  \bibinfo{author}{\bibfnamefont{J.}~\bibnamefont{McGreevy}}, \bibnamefont{and}
  \bibinfo{author}{\bibfnamefont{D.}~\bibnamefont{Vegh}},
  \bibinfo{journal}{Phys.Rev.} \textbf{\bibinfo{volume}{D83}},
  \bibinfo{pages}{065029} (\bibinfo{year}{2011}), \eprint{0903.2477}.

\bibitem[{\citenamefont{Faulkner
  et~al.}(2011{\natexlab{a}})\citenamefont{Faulkner, Iqbal, Liu, McGreevy, and
  Vegh}}]{Faulkner:2011tm}
\bibinfo{author}{\bibfnamefont{T.}~\bibnamefont{Faulkner}},
  \bibinfo{author}{\bibfnamefont{N.}~\bibnamefont{Iqbal}},
  \bibinfo{author}{\bibfnamefont{H.}~\bibnamefont{Liu}},
  \bibinfo{author}{\bibfnamefont{J.}~\bibnamefont{McGreevy}}, \bibnamefont{and}
  \bibinfo{author}{\bibfnamefont{D.}~\bibnamefont{Vegh}}
  (\bibinfo{year}{2011}{\natexlab{a}}), \bibinfo{note}{28 pages, many figures},
  \eprint{1101.0597}.

\bibitem[{\citenamefont{Faulkner and Polchinski}(2011)}]{Faulkner:2010tq}
\bibinfo{author}{\bibfnamefont{T.}~\bibnamefont{Faulkner}} \bibnamefont{and}
  \bibinfo{author}{\bibfnamefont{J.}~\bibnamefont{Polchinski}},
  \bibinfo{journal}{JHEP} \textbf{\bibinfo{volume}{1106}}, \bibinfo{pages}{012}
  (\bibinfo{year}{2011}), \eprint{1001.5049}.

\bibitem[{\citenamefont{Sachdev}(2010{\natexlab{b}})}]{Sachdev:2010um}
\bibinfo{author}{\bibfnamefont{S.}~\bibnamefont{Sachdev}},
  \bibinfo{journal}{Phys.Rev.Lett.} \textbf{\bibinfo{volume}{105}},
  \bibinfo{pages}{151602} (\bibinfo{year}{2010}{\natexlab{b}}),
  \eprint{1006.3794}.

\bibitem[{\citenamefont{Sachdev and Mueller}(2008)}]{Sachdev:2008ba}
\bibinfo{author}{\bibfnamefont{S.}~\bibnamefont{Sachdev}} \bibnamefont{and}
  \bibinfo{author}{\bibfnamefont{M.}~\bibnamefont{Mueller}}
  (\bibinfo{year}{2008}), \bibinfo{note}{talk at LT25, Amsterdam},
  \eprint{0810.3005}.

\bibitem[{\citenamefont{Faulkner
  et~al.}(2011{\natexlab{b}})\citenamefont{Faulkner, Liu, McGreevy, and
  Vegh}}]{Faulkner:2009wj}
\bibinfo{author}{\bibfnamefont{T.}~\bibnamefont{Faulkner}},
  \bibinfo{author}{\bibfnamefont{H.}~\bibnamefont{Liu}},
  \bibinfo{author}{\bibfnamefont{J.}~\bibnamefont{McGreevy}}, \bibnamefont{and}
  \bibinfo{author}{\bibfnamefont{D.}~\bibnamefont{Vegh}},
  \bibinfo{journal}{Phys.Rev.} \textbf{\bibinfo{volume}{D83}},
  \bibinfo{pages}{125002} (\bibinfo{year}{2011}{\natexlab{b}}),
  \eprint{0907.2694}.

\bibitem[{\citenamefont{Iqbal et~al.}(2010)\citenamefont{Iqbal, Liu, Mezei, and
  Si}}]{Iqbal:2010eh}
\bibinfo{author}{\bibfnamefont{N.}~\bibnamefont{Iqbal}},
  \bibinfo{author}{\bibfnamefont{H.}~\bibnamefont{Liu}},
  \bibinfo{author}{\bibfnamefont{M.}~\bibnamefont{Mezei}}, \bibnamefont{and}
  \bibinfo{author}{\bibfnamefont{Q.}~\bibnamefont{Si}},
  \bibinfo{journal}{Phys.Rev.} \textbf{\bibinfo{volume}{D82}},
  \bibinfo{pages}{045002} (\bibinfo{year}{2010}), \eprint{1003.0010}.

\bibitem[{\citenamefont{Iqbal et~al.}(2011{\natexlab{a}})\citenamefont{Iqbal,
  Liu, and Mezei}}]{Iqbal:2011ae}
\bibinfo{author}{\bibfnamefont{N.}~\bibnamefont{Iqbal}},
  \bibinfo{author}{\bibfnamefont{H.}~\bibnamefont{Liu}}, \bibnamefont{and}
  \bibinfo{author}{\bibfnamefont{M.}~\bibnamefont{Mezei}}
  (\bibinfo{year}{2011}{\natexlab{a}}), \bibinfo{note}{70 pages. Based on
  lectures given by Hong Liu}, \eprint{1110.3814}.

\bibitem[{\citenamefont{Iqbal et~al.}(2011{\natexlab{b}})\citenamefont{Iqbal,
  Liu, and Mezei}}]{Iqbal:2011aj}
\bibinfo{author}{\bibfnamefont{N.}~\bibnamefont{Iqbal}},
  \bibinfo{author}{\bibfnamefont{H.}~\bibnamefont{Liu}}, \bibnamefont{and}
  \bibinfo{author}{\bibfnamefont{M.}~\bibnamefont{Mezei}}
  (\bibinfo{year}{2011}{\natexlab{b}}), \eprint{1108.0425}.

\bibitem[{\citenamefont{Gubser and Pufu}(2008)}]{Gubser:2008wv}
\bibinfo{author}{\bibfnamefont{S.~S.} \bibnamefont{Gubser}} \bibnamefont{and}
  \bibinfo{author}{\bibfnamefont{S.~S.} \bibnamefont{Pufu}},
  \bibinfo{journal}{JHEP} \textbf{\bibinfo{volume}{0811}}, \bibinfo{pages}{033}
  (\bibinfo{year}{2008}), \eprint{0805.2960}.

\bibitem[{\citenamefont{Chen et~al.}(2010)\citenamefont{Chen, Kao, Maity, Wen,
  and Yeh}}]{Chen:2010mk}
\bibinfo{author}{\bibfnamefont{J.-W.} \bibnamefont{Chen}},
  \bibinfo{author}{\bibfnamefont{Y.-J.} \bibnamefont{Kao}},
  \bibinfo{author}{\bibfnamefont{D.}~\bibnamefont{Maity}},
  \bibinfo{author}{\bibfnamefont{W.-Y.} \bibnamefont{Wen}}, \bibnamefont{and}
  \bibinfo{author}{\bibfnamefont{C.-P.} \bibnamefont{Yeh}},
  \bibinfo{journal}{Phys.Rev.} \textbf{\bibinfo{volume}{D81}},
  \bibinfo{pages}{106008} (\bibinfo{year}{2010}), \eprint{1003.2991}.

\bibitem[{\citenamefont{Gubser}(2008)}]{Gubser:2008px}
\bibinfo{author}{\bibfnamefont{S.~S.} \bibnamefont{Gubser}},
  \bibinfo{journal}{Phys.Rev.} \textbf{\bibinfo{volume}{D78}},
  \bibinfo{pages}{065034} (\bibinfo{year}{2008}), \eprint{0801.2977}.

\bibitem[{\citenamefont{Herzog et~al.}(2007)\citenamefont{Herzog, Kovtun,
  Sachdev, and Son}}]{Herzog:2007ij}
\bibinfo{author}{\bibfnamefont{C.~P.} \bibnamefont{Herzog}},
  \bibinfo{author}{\bibfnamefont{P.}~\bibnamefont{Kovtun}},
  \bibinfo{author}{\bibfnamefont{S.}~\bibnamefont{Sachdev}}, \bibnamefont{and}
  \bibinfo{author}{\bibfnamefont{D.~T.} \bibnamefont{Son}},
  \bibinfo{journal}{Phys.Rev.} \textbf{\bibinfo{volume}{D75}},
  \bibinfo{pages}{085020} (\bibinfo{year}{2007}), \eprint{hep-th/0701036}.

\bibitem[{\citenamefont{Son and Starinets}(2002)}]{Son:2002sd}
\bibinfo{author}{\bibfnamefont{D.~T.} \bibnamefont{Son}} \bibnamefont{and}
  \bibinfo{author}{\bibfnamefont{A.~O.} \bibnamefont{Starinets}},
  \bibinfo{journal}{JHEP} \textbf{\bibinfo{volume}{0209}}, \bibinfo{pages}{042}
  (\bibinfo{year}{2002}), \eprint{hep-th/0205051}.

\bibitem[{\citenamefont{{Olver}}(1954)}]{1954RSPTA.247..307O}
\bibinfo{author}{\bibfnamefont{F.~W.~J.} \bibnamefont{{Olver}}},
  \bibinfo{journal}{Royal Society of London Philosophical Transactions Series
  A} \textbf{\bibinfo{volume}{247}}, \bibinfo{pages}{307}
  (\bibinfo{year}{1954}).

\bibitem[{\citenamefont{Policastro and Starinets}(2001)}]{Policastro:2001yb}
\bibinfo{author}{\bibfnamefont{G.}~\bibnamefont{Policastro}} \bibnamefont{and}
  \bibinfo{author}{\bibfnamefont{A.}~\bibnamefont{Starinets}},
  \bibinfo{journal}{Nucl.Phys.} \textbf{\bibinfo{volume}{B610}},
  \bibinfo{pages}{117} (\bibinfo{year}{2001}), \eprint{hep-th/0104065}.

\bibitem[{\citenamefont{N\'u\~nez and Starinets}(2003)}]{PhysRevD.67.124013}
\bibinfo{author}{\bibfnamefont{A.}~\bibnamefont{N\'u\~nez}} \bibnamefont{and}
  \bibinfo{author}{\bibfnamefont{A.~O.} \bibnamefont{Starinets}},
  \bibinfo{journal}{Phys. Rev. D} \textbf{\bibinfo{volume}{67}},
  \bibinfo{pages}{124013} (\bibinfo{year}{2003}).

\bibitem[{\citenamefont{{Luo}}(2011)}]{2011JHEP...11..142L}
\bibinfo{author}{\bibfnamefont{M.~J.} \bibnamefont{{Luo}}},
  \bibinfo{journal}{Journal of High Energy Physics}
  \textbf{\bibinfo{volume}{11}}, \bibinfo{pages}{142} (\bibinfo{year}{2011}),
  \eprint{1111.3992}.

\bibitem[{\citenamefont{Keimer et~al.}(1991)\citenamefont{Keimer, Birgeneau,
  Cassanho, Endoh, Erwin, Kastner, and Shirane}}]{PhysRevLett.67.1930}
\bibinfo{author}{\bibfnamefont{B.}~\bibnamefont{Keimer}},
  \bibinfo{author}{\bibfnamefont{R.~J.} \bibnamefont{Birgeneau}},
  \bibinfo{author}{\bibfnamefont{A.}~\bibnamefont{Cassanho}},
  \bibinfo{author}{\bibfnamefont{Y.}~\bibnamefont{Endoh}},
  \bibinfo{author}{\bibfnamefont{R.~W.} \bibnamefont{Erwin}},
  \bibinfo{author}{\bibfnamefont{M.~A.} \bibnamefont{Kastner}},
  \bibnamefont{and} \bibinfo{author}{\bibfnamefont{G.}~\bibnamefont{Shirane}},
  \bibinfo{journal}{Phys. Rev. Lett.} \textbf{\bibinfo{volume}{67}},
  \bibinfo{pages}{1930} (\bibinfo{year}{1991}).

\bibitem[{\citenamefont{Varma et~al.}(1989)\citenamefont{Varma, Littlewood,
  Schmitt-Rink, Abrahams, and Ruckenstein}}]{PhysRevLett.63.1996}
\bibinfo{author}{\bibfnamefont{C.~M.} \bibnamefont{Varma}},
  \bibinfo{author}{\bibfnamefont{P.~B.} \bibnamefont{Littlewood}},
  \bibinfo{author}{\bibfnamefont{S.}~\bibnamefont{Schmitt-Rink}},
  \bibinfo{author}{\bibfnamefont{E.}~\bibnamefont{Abrahams}}, \bibnamefont{and}
  \bibinfo{author}{\bibfnamefont{A.~E.} \bibnamefont{Ruckenstein}},
  \bibinfo{journal}{Phys. Rev. Lett.} \textbf{\bibinfo{volume}{63}},
  \bibinfo{pages}{1996} (\bibinfo{year}{1989}).

\bibitem[{\citenamefont{Gegenwart et~al.}(2005)\citenamefont{Gegenwart,
  Custers, Tokiwa, Geibel, and Steglich}}]{PhysRevLett.94.076402}
\bibinfo{author}{\bibfnamefont{P.}~\bibnamefont{Gegenwart}},
  \bibinfo{author}{\bibfnamefont{J.}~\bibnamefont{Custers}},
  \bibinfo{author}{\bibfnamefont{Y.}~\bibnamefont{Tokiwa}},
  \bibinfo{author}{\bibfnamefont{C.}~\bibnamefont{Geibel}}, \bibnamefont{and}
  \bibinfo{author}{\bibfnamefont{F.}~\bibnamefont{Steglich}},
  \bibinfo{journal}{Phys. Rev. Lett.} \textbf{\bibinfo{volume}{94}},
  \bibinfo{pages}{076402} (\bibinfo{year}{2005}).

\bibitem[{\citenamefont{Helton et~al.}(2010)\citenamefont{Helton, Matan,
  Shores, Nytko, Bartlett, Qiu, Nocera, and Lee}}]{PhysRevLett.104.147201}
\bibinfo{author}{\bibfnamefont{J.~S.} \bibnamefont{Helton}},
  \bibinfo{author}{\bibfnamefont{K.}~\bibnamefont{Matan}},
  \bibinfo{author}{\bibfnamefont{M.~P.} \bibnamefont{Shores}},
  \bibinfo{author}{\bibfnamefont{E.~A.} \bibnamefont{Nytko}},
  \bibinfo{author}{\bibfnamefont{B.~M.} \bibnamefont{Bartlett}},
  \bibinfo{author}{\bibfnamefont{Y.}~\bibnamefont{Qiu}},
  \bibinfo{author}{\bibfnamefont{D.~G.} \bibnamefont{Nocera}},
  \bibnamefont{and} \bibinfo{author}{\bibfnamefont{Y.~S.} \bibnamefont{Lee}},
  \bibinfo{journal}{Phys. Rev. Lett.} \textbf{\bibinfo{volume}{104}},
  \bibinfo{pages}{147201} (\bibinfo{year}{2010}).

\bibitem[{\citenamefont{Aronson et~al.}(1995)\citenamefont{Aronson, Osborn,
  Robinson, Lynn, Chau, Seaman, and Maple}}]{PhysRevLett.75.725}
\bibinfo{author}{\bibfnamefont{M.~C.} \bibnamefont{Aronson}},
  \bibinfo{author}{\bibfnamefont{R.}~\bibnamefont{Osborn}},
  \bibinfo{author}{\bibfnamefont{R.~A.} \bibnamefont{Robinson}},
  \bibinfo{author}{\bibfnamefont{J.~W.} \bibnamefont{Lynn}},
  \bibinfo{author}{\bibfnamefont{R.}~\bibnamefont{Chau}},
  \bibinfo{author}{\bibfnamefont{C.~L.} \bibnamefont{Seaman}},
  \bibnamefont{and} \bibinfo{author}{\bibfnamefont{M.~B.} \bibnamefont{Maple}},
  \bibinfo{journal}{Phys. Rev. Lett.} \textbf{\bibinfo{volume}{75}},
  \bibinfo{pages}{725} (\bibinfo{year}{1995}).

\bibitem[{\citenamefont{P\'epin}(2005)}]{PhysRevLett.94.066402}
\bibinfo{author}{\bibfnamefont{C.}~\bibnamefont{P\'epin}},
  \bibinfo{journal}{Phys. Rev. Lett.} \textbf{\bibinfo{volume}{94}},
  \bibinfo{pages}{066402} (\bibinfo{year}{2005}).

\bibitem[{\citenamefont{{Si} et~al.}(2001)\citenamefont{{Si}, {Rabello},
  {Ingersent}, and {Smith}}}]{2001Natur.413..804S}
\bibinfo{author}{\bibfnamefont{Q.}~\bibnamefont{{Si}}},
  \bibinfo{author}{\bibfnamefont{S.}~\bibnamefont{{Rabello}}},
  \bibinfo{author}{\bibfnamefont{K.}~\bibnamefont{{Ingersent}}},
  \bibnamefont{and} \bibinfo{author}{\bibfnamefont{J.~L.}
  \bibnamefont{{Smith}}}, \bibinfo{journal}{\nat}
  \textbf{\bibinfo{volume}{413}}, \bibinfo{pages}{804} (\bibinfo{year}{2001}),
  \eprint{arXiv:cond-mat/0011477}.

\bibitem[{\citenamefont{{Abrahams} and {Varma}}(2000)}]{2000PNAS...97.5714A}
\bibinfo{author}{\bibfnamefont{E.}~\bibnamefont{{Abrahams}}} \bibnamefont{and}
  \bibinfo{author}{\bibfnamefont{C.~M.} \bibnamefont{{Varma}}},
  \bibinfo{journal}{Proceedings of the National Academy of Science}
  \textbf{\bibinfo{volume}{97}}, \bibinfo{pages}{5714} (\bibinfo{year}{2000}),
  \eprint{arXiv:cond-mat/0003135}.

\end{thebibliography}

\end{document}